# Composition-dependent polarization switching behaviors of (111)-preferred polycrystalline Pb(Zr$_x$Ti$_{1-x}$)O$_3$ thin films

J. Y. Jo, S. M. Yang, H. S. Han, D. J. Kim, W. S. Choi, and T. W. Noh[a]

*ReCOE & FPRD, Department of Physics and Astronomy, Seoul National University, Seoul 151-747, Korea*

T. K. Song

*School of Nano and Advanced Materials Engineering, Changwon National University, Changwon, Gyeongnam 641-773, Korea*

J.-G. Yoon

*Department of Physics, University of Suwon, Gyeonggi-do 445-743, Korea*

C.-Y. Koo, J.-H. Cheon, and S.-H. Kim

*R&D center, INOSTEK Inc., Ansan, Gyeonggi-do 426-901, Korea*

[a] E-mail: twnoh@snu.ac.kr

We investigated the time-dependent polarization switching behaviors of (111)-preferred polycrystalline Pb(Zr$_x$Ti$_{1-x}$)O$_3$ thin films with various Zr concentrations. We could explain all the polarization switching behaviors well by assuming Lorentzian distributions in the logarithmic polarization switching time [Refer to J. Y. Jo *et al*., Phys. Rev. Lett. (in press)]. Based on this analysis, we found that the Zr ion-substitution for Ti ions would induce broad distributions in the local field due to defect dipoles, which makes the ferroelectric domain switching occur more easily.

Pb($Zr_x Ti_{1-x}$)$O_3$ (PZT) film is a ferroelectric (FE) material which is widely used in numerous applications, including ferroelectric random access memories (FeRAM).[1] The replacement of Ti with Zr ions in $PbTiO_3$ causes marked changes in the FE properties, such as the transition temperature, the remnant polarization, and the coercive field. Further, the associated FE domain switching dynamics would be expected to change significantly. For example, many workers have reported that the polarization-voltage (P-V) hysteresis loops of PZT films change from an almost square shape to a slanted shape with increase of $x$.[2,3] Despite its importance in actual applications, the effects of Zr variation on FE domain switching dynamics of PZT films have rarely been studied.

The Kolmogorov-Avrami-Ishibashi model[4,5,6] has been used successfully to describe FE domain switching kinetics in single crystals and epitaxial films.[7] According to this classical model, time($t$)-dependent changes in polarization can be written as $\Delta P_{KAI}(t) = 2P_s[1-exp\{-(t/t_0)^n\}]$, where $P_s$ is the spontaneous polarization, and $n$ and $t_0$ are the effective dimension and characteristic switching time of the domain growth, respectively. Recently, we investigated domain switching dynamics of (111)-preferred polycrystalline $PbZr_{0.3}Ti_{0.7}O_3$ films.[8] We showed that the $t$-dependent change in polarization, $\Delta P(t)$, for the polycrystalline PZT films can be well described by

$$\Delta P(t) = \int_{-\infty}^{\infty} \Delta P_{KAI} F(log\ t_0) d(log\ t_0), \quad (1)$$

where $F(log\ t_0)$ can be written as a Lorentzian function:

$$F(log\ t_0) = \frac{A}{\pi} \cdot \frac{w}{(log\ t_0 - log\ t_1)^2 + w^2}. \quad (2)$$

We proposed that the Lorentzian function results from local field variations caused by dipole defects at domain pinning sites.[8]

In this Letter, we report the domain switching dynamics of (111)-preferred polycrystalline Pb($Zr_x Ti_{1-x}$)$O_3$ thin films at various values of $x$. We found that all our experimental $\Delta P(t)$ data can be fitted very well using Lorentzian functions for $F(log\ t_0)$. With increase of $x$, $F(log\ t_0)$ broadens, indicating the increase of the local field variation. We compare the details of our studies with earlier work, including electron paramagnetic resonance studies.[9,10,11] Further, our work can also explain the systematic changes in *P-V* hysteresis loops with variations in *x*.

We prepared (111)-preferred polycrystalline PZT thin films with *x* values of 0.30, 0.35, 0.40, and 0.52. We deposited polycrystalline PZT films on Pt/Ti/SiO$_2$/Si substrates using chemical solution deposition method. The PZT film thicknesses were about 150 nm. Then, we deposited Pt top electrodes with area of $7.9 \times 10^{-9}$ m$^2$ using sputtering with a shadow mask. X-ray diffraction studies showed that the PZT films were all in the (111)-preferred orientation, as shown in Fig. 1(a). All *P-V* hysteresis loops measured at 2 kHz triangular waves (aixACCT TF Analyzer2000) showed ferroelectric responses, as displayed in Fig. 1(b). With increase of *x*, the *P-V* hysteresis loops changed shape from a nearly square to a slanted one.[2,3]

We measured $\Delta P(t)$ of the PZT thin films using pulse measurements, as described previously.[8] Figures 2(a) and (b) show schematic diagrams of the pulse trains used to measure the non-switching and switching polarizations, respectively. Their difference corresponds to $\Delta P(t)$. Writing pulse widths (*t*) varied from 200 ns to 1 ms, and the pulse heights ($V_{ext}$) varied from 0.5 to 2.5 V. We could estimate the value of external electric field, $E_{ext}$, by dividing $V_{ext}$ by the film thickness.

Figure 2(c) shows the normalized time-dependent polarization change, $\Delta P(t)/2P_s$, at room temperature for the PZT film with *x* = 0.30. With increase of $V_{ext}$, polarization switching occurred earlier. Note that polarization switching can occur for $V_{ext} < V_C$, whereas the coercive voltage, $V_C$, of the film, is about 1.0 V. Figures 2(c)-(f) show the *x* dependence of the domain switching dynamics. With increase of *x*, differences between the $\Delta P(t)/2P_s$ values for $V_{ext}$ = 0.8 and 1.5 V become smaller for any given *t*. This *x*-dependence is consistent with the observed systematic slope changes in the *P-V* hysteresis loops, shown in Fig. 1(b). It should be noted that the $\Delta P(t)/2P_s$ data provide much more information on domain switching dynamics than simple *P-V* measurements at a given frequency.

We could fit all of the measured $\Delta P(t)/2P_s$ data quite well using Eqs. (1) and (2). The solid lines in Figs. 2(c)-(f) show the best results. For example, Figure 2(g) shows the $F(\log t_0)$ used to fit data at *x* = 0.30. With increase of $V_{ext}$, the center of the Lorentzian distribution, i.e., $\log t_1$ in Eq. (2), decreases significantly. In particular, for the *x* = 0.30 sample, $t_1$ varied by about 4 orders of magnitude as $V_{ext}$ increased by a factor of 2.

Figures 2(g)-(j) show $F(log\ t_0)$ used for fitting other data. The good agreements between the experimental data and the theoretical fits suggest the possible universality of the Lorentzian distribution for $F(log\ t_0)$ for all (111)-preferred PZT films.

Recently, we proposed that the Lorentzian function for $F(log\ t_0)$ could be closely related to dipole defects inside FE materials and hence used to obtain related microscopic information.[8] Note that the dipole defects inside FE materials will act as pinning sites for domain wall motion.[12] By using Lévy distribution arguments,[13] we showed that the local field $\bar{E}$, due to randomly distributed dipole defects, should result in a Lorentzian function for $F(\bar{E})$,[8] namely:

$$F(\bar{E}) = \frac{A}{\pi} \cdot \frac{\Delta}{\bar{E}^2 + \Delta^2} . \quad (3)$$

In the low $E_{ext}$ region, which is known as the "creep motion" region,[14] the domain wall motion should be governed by thermal activation processes at the pinning sites. Applying Arrhenius's law and Eqs. (2) and (3),[8,15] $log\ t_1$ and $w$ can be related to the microscopic quantities:[8]

$$log\ t_1 \approx \frac{\alpha}{E_{ext}}, \quad (4)$$

and

$$w \approx \Delta \cdot \frac{\alpha}{E_{ext}^2}, \quad (5)$$

where $\alpha$ and $\Delta$ are the activation field for domain wall motion and the half-width at half-maximum of the local field distribution $F(\bar{E})$, respectively.

Figure 3(a) shows a plot of ($log\ t_1$) vs. ($1/E_{ext}$) for PZT films with various $x$ values. With increase of $1/E_{ext}$, $log\ t_1$ increases. In the low $E_{ext}$ region, $log\ t_1$ values fall into linear lines. From these slopes, we estimated the $\alpha$ values using Eq. (4). The solid (blue) squares in Fig. 3(c) show that $\alpha$ decreases with increase of $x$. That is, domain switching can occur more easily at higher Zr concentrations.

Figure 3(b) shows the plot of ($w$) vs. ($\alpha/E_{ext}^2$) for the PZT films. As $1/E_{ext}^2$ increases, $w$ increases. In the low $E_{ext}$ region, we obtained $\Delta$ values using Eq.(5). These represent the broadening of the local field due to the dipole defects inside the PZT films. The solid (red) circles in Fig. 3(c) show that $\Delta$ increases with increase of $x$ and that the local field has a broader distribution at higher Zr concentration.

Our Zr concentration-dependence studies provide a physical understanding on how the FE domain dynamics of polycrystalline PZT films vary. Note that this understanding of the microscopic processes is in good agreement with those obtained from other experimental work. In recent microscopic studies of defect dipoles in bulk PbZr$_x$Ti$_{1-x}$O$_3$, using electron paramagnetic resonance.[9,10,11] The resonance peaks broadened similarly to our $F(log\ t_0)$. The authors explained qualitatively that the Zr ions induced a broader angular distribution of defect dipoles with increase of $x$. Their results were consistent with the broadening of local field distribution, displayed in Fig. 3(c). It should be noted that our studies have not only provided an explanation on the role of the defect dipoles in FE domain wall motion, but also quantitative values for $\alpha$ and $\Delta$ at various values of $x$.

Finally, let us go back to the $x$-dependent systematic changes of the $P$-$V$ hysteresis loop, observed in Fig. 1(b). The effective electric field for domain wall motion for a given site should be the sum of $E_{ext}$ and $\bar{E}$. Polarization switching can then occur for $V_{ext} < V_C$, as shown in Fig. 2(c). With a larger value of $x$, the larger variation in $\bar{E}$ should result in a smaller polarization change for a given $E_{ext}$ variation. Therefore, the $P$-$V$ hysteresis loop for the sample with a larger $x$ value should become slanted.

In summary, we found that the Zr composition of our (111)-preferred PbZr$_x$Ti$_{1-x}$O$_3$ films could strongly affect polarization switching dynamics. With increase of $x$, the local field due to defect dipoles has a broader distribution and domain switching can occur easily.

We thank S.-H. Bae and Chan Park for their help with the x-ray diffraction experiments and Y. H. Oh and Y.-W. Kim for discussions. This study was financially supported by the Creative Research Initiative program (Functionally Integrated Oxide Heterostructure) of MOST/KOSEF.

**Figure Legends**

Fig. 1 (Color online) (a) X-ray diffraction results and (b) *P-V* hysteresis loops of the (111)-preferred Pb(Zr$_x$Ti$_{1-x}$)O$_3$ films.

Fig. 2 (Color online) Schematic diagrams of the pulse trains used to measure (a) non-switching polarization, *P*$_{ns}$, and (b) switching polarization, *P*$_{sw}$. Plots (c)-(f) show time (*t*)-dependent polarization changes $\Delta P(t)$ of Pb(Zr$_x$Ti$_{1-x}$)O$_3$ films under various external voltages, *V*$_{ext}$. Plots (g)-(j) show the corresponding distribution functions of the logarithmic characteristic switching time $t_0$, *F*(*log t*$_0$).

Fig. 3 (Color online) External electric field (*E*$_{ext}$)-dependent changes of parameters in *F*(*log t*$_0$). (a) The center value and (b) the width of the distribution. (c) Variations of activation fields, *α* (solid blue squares), and half width, *Δ*, at half maximum of local field distributions (solid red circles) with a Zr concentration change.

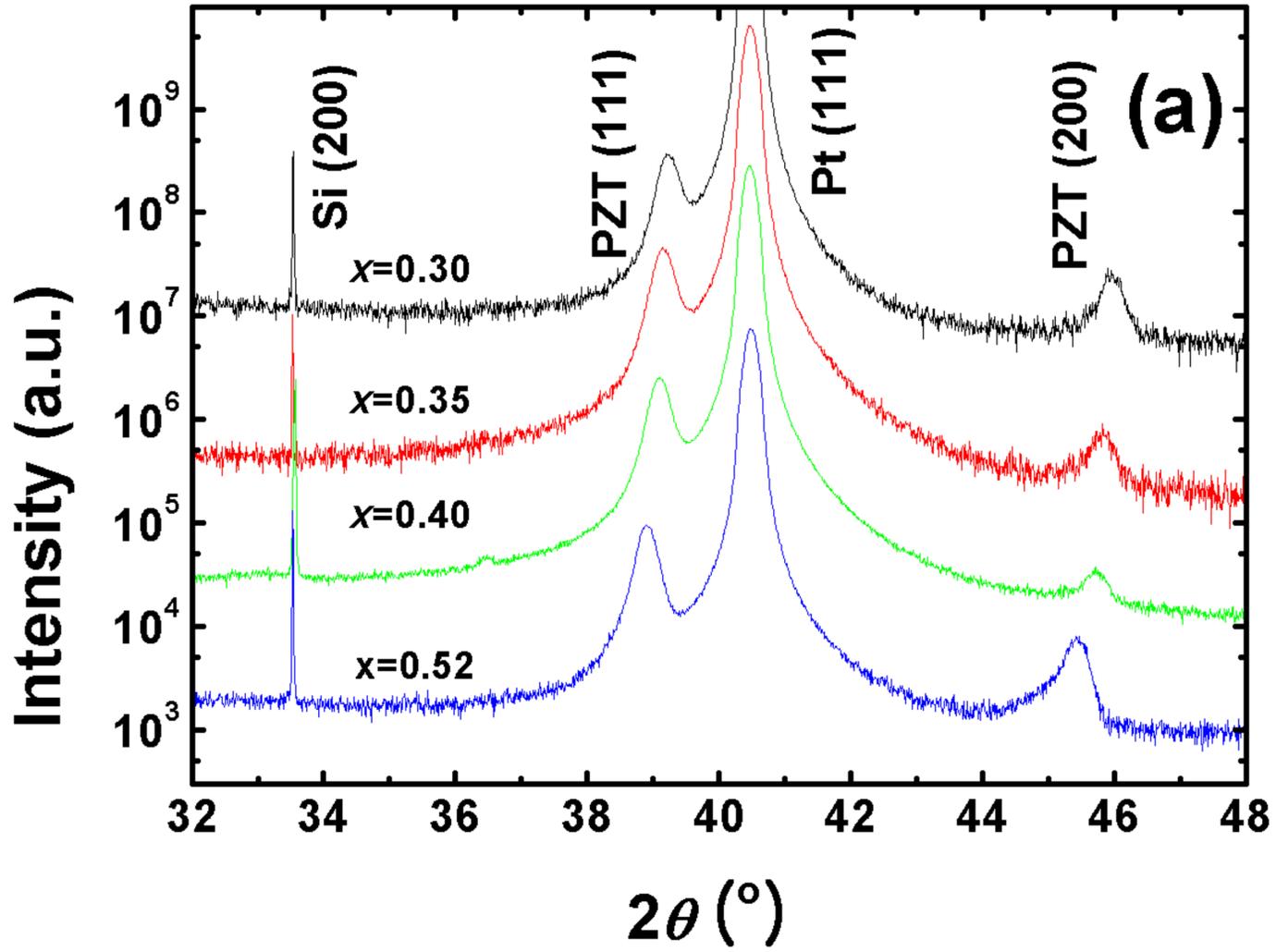

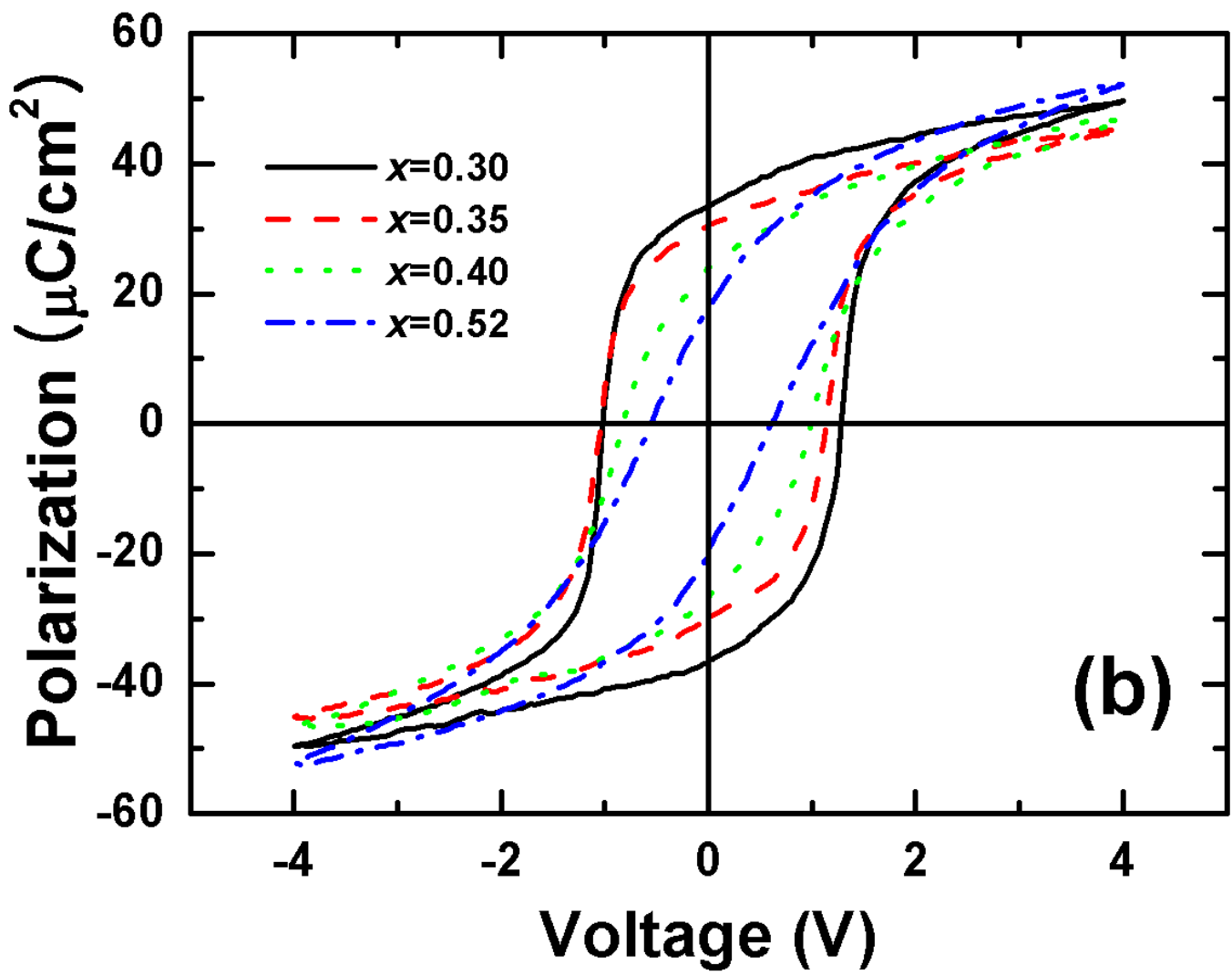

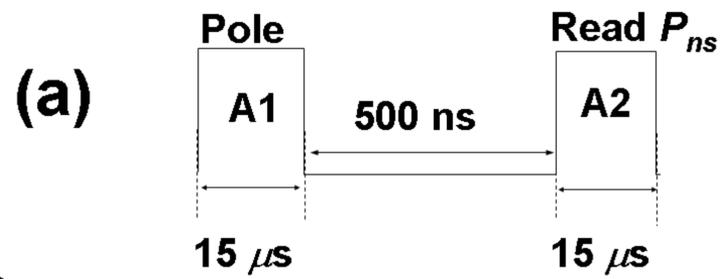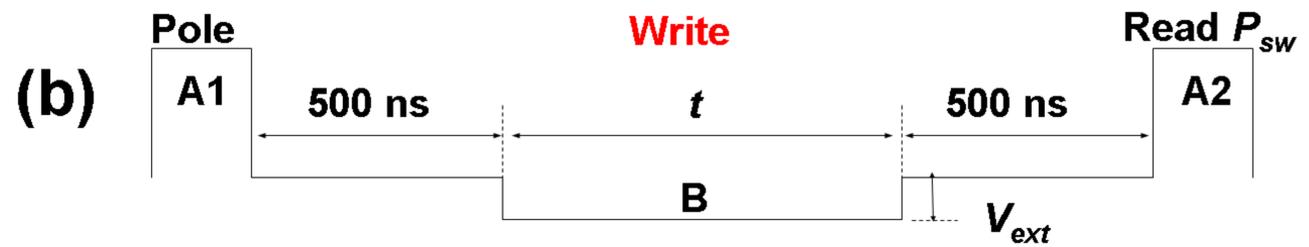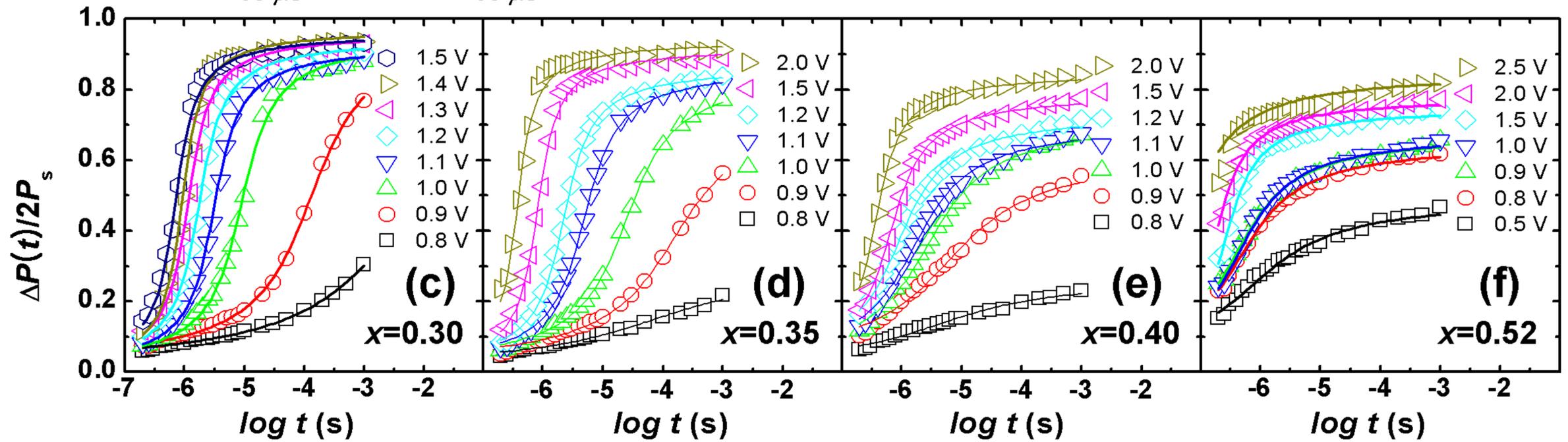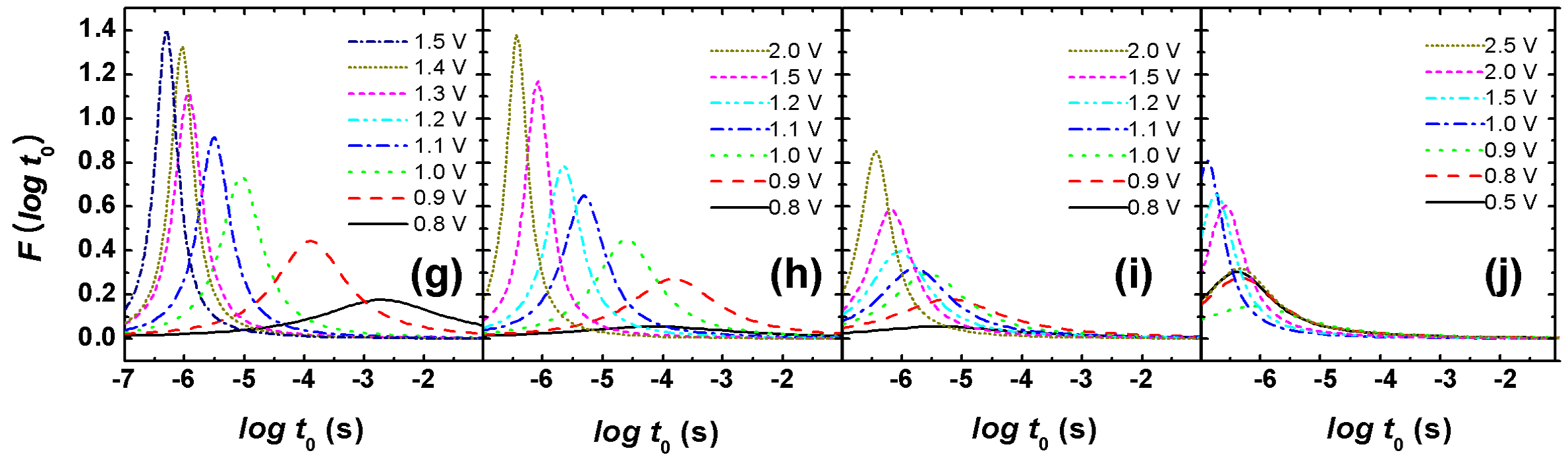

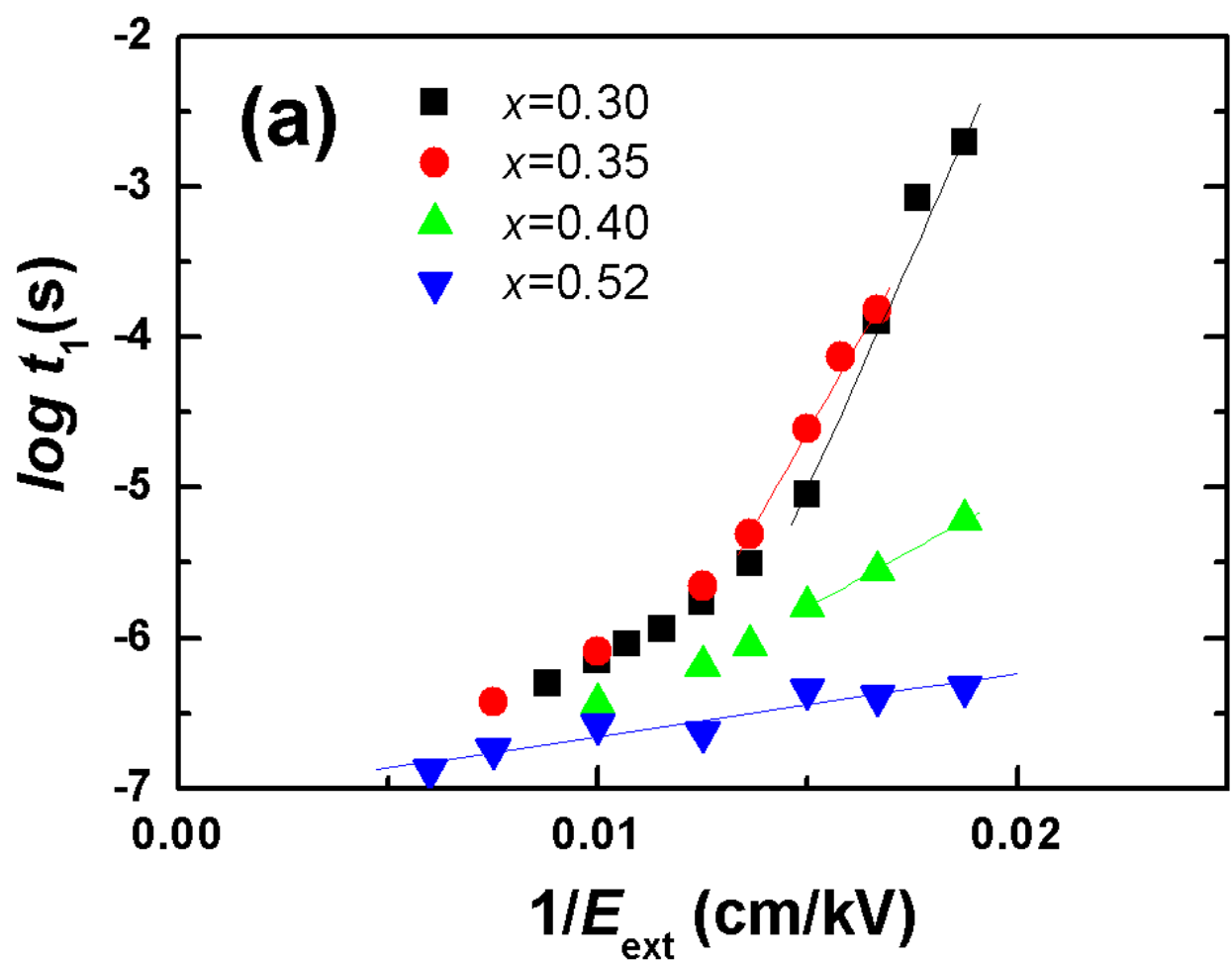
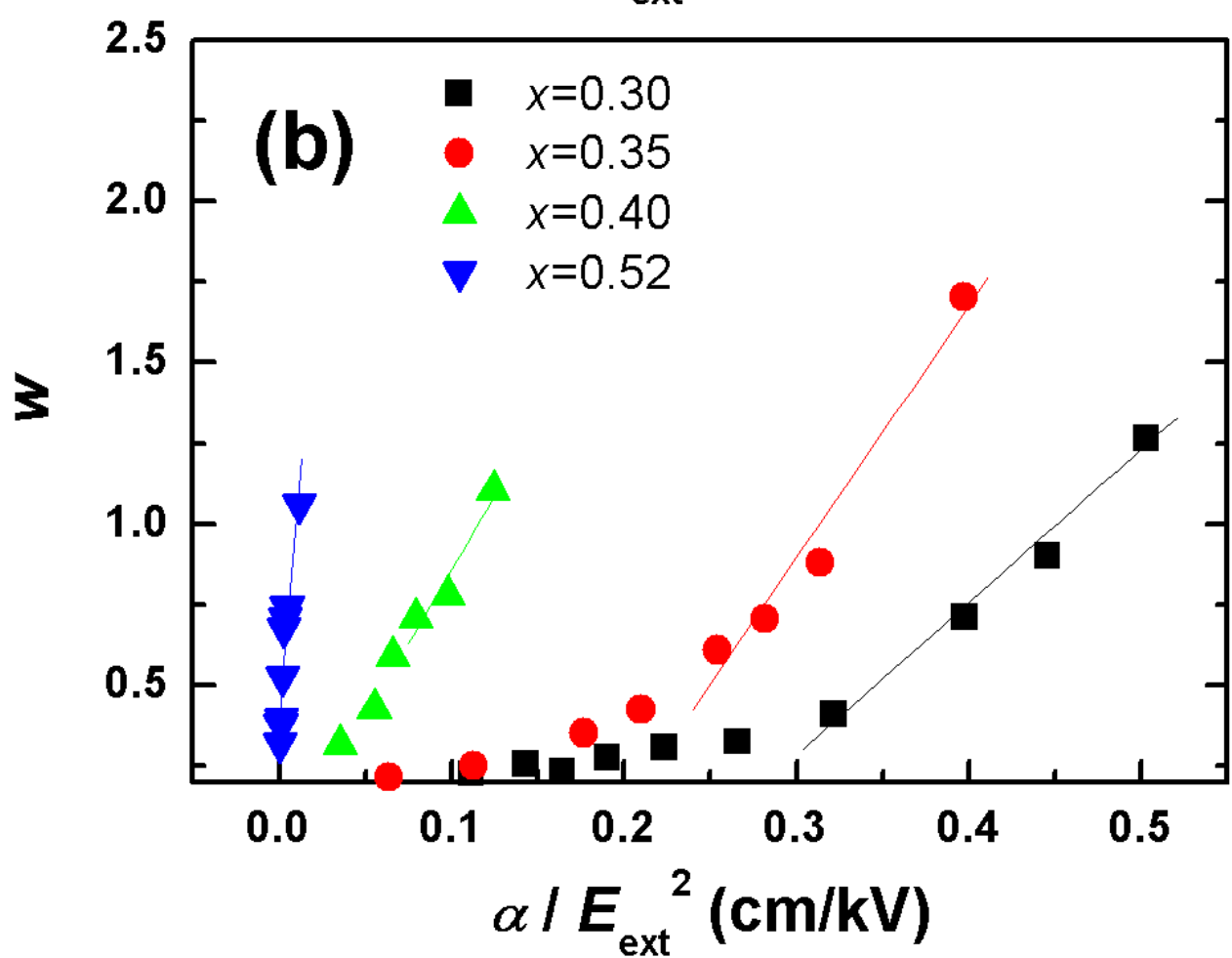
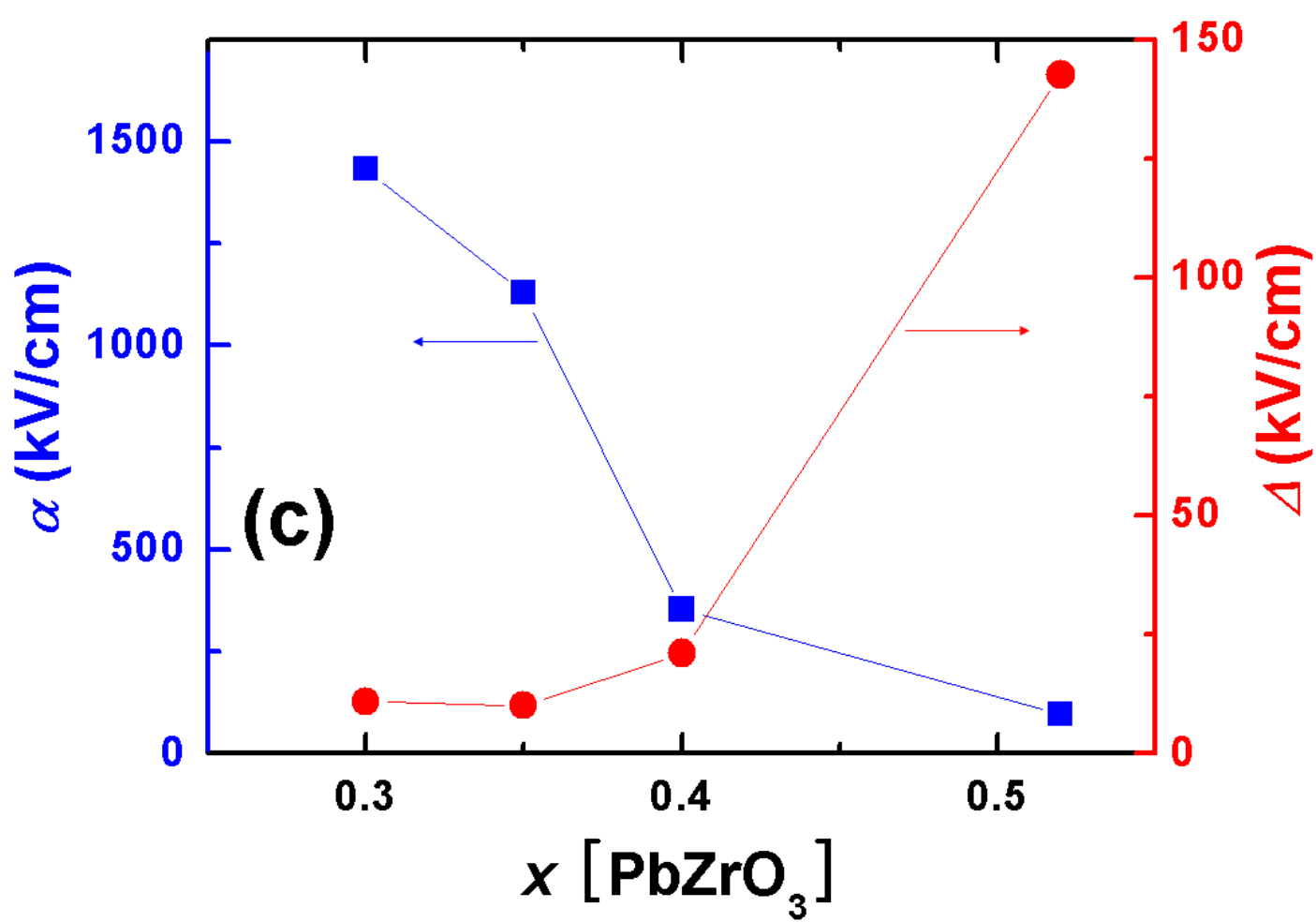